\documentstyle[12pt]{article}
\newcommand{\doublespace}{
    \renewcommand{\baselinestretch}{1.6}\large\normalsize}

\newcommand{\be}{\begin{equation}}
\newcommand{\ee}{\end{equation}}
\newcommand{\ba}{\begin{eqnarray}}
\newcommand{\ea}{\end{eqnarray}}

\newcommand{\ave}[1]{\langle {#1} \rangle}

\begin{document}
\begin{titlepage}
\pagestyle{empty}
\vspace{1.0in}
\begin{flushright}
SUNY-NTG-94-61
\end{flushright}
\begin{flushright}
December 1994
\end{flushright}
\vspace{1.0in}
\begin{center}
\doublespace
\begin{large}
{\bf{ROLE OF CLOUD RENORMALIZATION IN CONVOLUTION MODELS FOR EMC AND DRELL-YAN
RATIOS}}\\
\end{large}
\vskip 1.0in
Michael Buballa\\
{\small
{\it Department of Physics, State University of New York,\\
Stony Brook, New York 11794, U.S.A.}}\\
\end{center}
\vspace{2cm}

\begin{abstract}
Generalizing a recent convolution model of the nucleon
which explicitly conserves charge and momentum we
reexamine the influence of an enhanced pion field in nuclei on EMC and
Drell-Yan ratios. Due to wave function renormalization constants the effect
is more than 50\% smaller than predicted by the standard pion excess model.
In particular there is no discrepancy between the EMC data and our results
in the $x$-region which was expected to be most sensitive to the pion.
\end{abstract}
\end{titlepage}

\doublespace

\setcounter{equation}{0}

When in 1983 the EMC collaboration found an enhanced nucleon structure
function $F_2(x)$ in iron at $x \leq 0.3$ \cite{emc},
this was interpreted as a signature for an enhanced pion field in nuclei which
was predicted already in the 70s \cite{edam}. Starting from
the pion distribution function of the free nucleon, calculated in 1974
by Sullivan \cite{sjdu}, Llewellyn Smith
\cite{llsm} was able to explain the effect by the assumption that the iron
nucleus contains about 0.15 additional pions per nucleon.
Generalizing Sullivan's expression Ericson and Thomas
\cite{etma} showed that these extra pions result in a rather natural way from
the attractive pion-nucleon interaction in a conventional meson-exchange
description of nuclear matter. Consequently, after the enhancement of $F_2$
could not be confirmed by later experiments \cite{apma,argr}, it was concluded
that there are no excess pions in nuclei, calling into question the
meson-exchange picture itself \cite{bfsg}. These conclusions were corroborated
by the non-existence of a nuclear dependence of Drell-Yan cross sections
\cite{adml}. Recent attempts to solve the problem are based on a possible
change of the gluon properties in nuclei \cite{bfsg} or the partial restoration
of chiral symmetry with density \cite{bblw}.
In this communication we do not intend to argue in favor or against these
suggestions. Instead we want to reinvestigate the question whether an enhanced
pion field -- if it exists -- really leads to a detectable enhancement of the
EMC and Drell-Yan ratios. In particular we want to discuss the role of
wavefunction renormalization constants which have not been considered in the
past.

Starting point of our investigations is the two-phase model for the nucleon
developed by Szczurek and Speth \cite{ss93}.
In this model the nucleon consists of a bare core and a meson-baryon phase.
Here we want to restrict ourselves to pion and nucleon degrees of freedom
which are supposed to be the main players in the region $0.1 \leq x \leq 0.3$.
Then the distribution function for a quark with flavor $f$ in a proton reads:
\be
   q_f^p(x) \;=\; Z_N \{ q_{f,bare}^p(x)
   \;+\; \sum_i c_i \;[\; \int_x^1 \frac{dy}{y} \; f^{N/N}(y) \;
   q_{f,bare}^{N_i} (\frac{x}{y})
   \;+\;  \int_x^1 \frac{dy}{y} \; f^{\pi/N}(y) \;
   q_f^{\pi_i} (\frac{x}{y})\;]\;\}\;.
\ee
For the sake of readability we do not write the $Q^2$-dependence of the
distribution functions explicitly.
The two convolution integrals are due to the pion-nucleon phase or, more
precisely, a $\pi^+ n$ phase with $c_i = 2/3$ and a
$\pi^o p$ phase with $c_i = 1/3$.
The function $f^{\pi/N}$ is the Sullivan expression for the momentum
distribution of the pion while $f^{N/N}$ is the analogous distribution function
of the recoil nucleon.
The sum is multiplied by an overall constant $Z_N$
which normalizes the total probability of finding the dressed nucleon
either in the bare nucleon phase or in the pion-nucleon phase to unity.
It is given by
\be
   Z_N^{-1} - 1 \;=\; M_1^{\pi/N} \equiv \int_0^1 dy\;f^{\pi/N}(y) \;=\;
   \int_0^1 dy\;f^{N/N}(y) \;.
\ee
This is a consequence
of the well-known number sum rules for valence quarks reflecting the
conservation of baryon number, charge and strangeness:
\be
\int_0^1 dx \;(q_f^p(x) - \bar q_f^p(x)) \;=\;
\int_0^1 dx \;(q_{f,bare}^p(x) - \bar q_{f,bare}^p(x)) \;=\;N^p_f \;,
\ee
with $N^p_u = 2$, $N^p_d = 1$ and $N^p_s = 0$. Note that eq.~(2) also
implies that the first moments of the distribution functions $f^{\pi/N}$
and $f^{N/N}$ have to be equal.

Similarly momentum conservation
\be
\sum_f \int_0^1 dx \;x\,q_f^p(x) \;=\; \sum_f\int_0^1 dx \;x\,q_{f,bare}^p(x)
\ee
leads to a relation between the second moments of $f^{\pi/N}$ and $f^{N/N}$
\cite{meth}. In eq.~(4) the flavor index $f$ runs over quarks and antiquarks.
In principle gluons have to be included, too. However, since experimentally
the momentum fraction carried by quarks is about the same in a pion as
it is in a nucleon ($\sim 47\%$ at $Q^2 = 50 (GeV/c)^2$) we must
demand momentum conservation for quarks and gluons separately when we want to
keep the internal structure of the constituents unchanged.

Higher moments of $f^{N/N}$ are not related to $f^{\pi/N}$ by conservation
laws. However,
according to the intuitive picture that the nucleon breaks up into a pion
which carries the fraction $y$ of the plus momentum of the original nucleon
and a recoil nucleon which carries the missing fraction $1-y$ it is frequently
assumed \cite{ss93}
\be
   f^{N/N}(y) \;=\; f^{\pi/N}(1-y).
\ee
This automatically guarantees the validity of eqs. (2) - (4) but is not
a necessary consequence of these relations.

Many features we want to
discuss later on can be expressed in terms of the mean number of pions per
nucleon, $\ave{n_\pi}$. For the free nucleon we get from eqs.~(1) and (2):
\be
   \ave{n_\pi^N} \;=\; Z_N\,M_1^{\pi/N} \;=\; \frac{M_1^{\pi/N}}{1 \,+\,
   M_1^{\pi/N}} \;.
\ee
Note that even for large moments $\ave{n_\pi}$ is always less than
one, in agreement with the two-phase picture of the nucleon we have started
with: Since at any time there is at most one pion, $\ave{n_\pi}$ should never
exceed unity. (Of course, for $M_1^\pi > 1$, i.e.
when the $N\pi$ loop becomes more important than the bare nucleon phase the
two-phase model itself becomes unrealistic and multiple pion phases have
to be included. In practice, however, this does not happen.)

The essential point of eq.~(1) is that the renormalization constant $Z_N$ is
an {\it overall} prefactor, i.e. not only the bare nucleon part but also the
cloud contribution becomes renormalized.
In contrast Melnitchouk and Thomas \cite{meth} argue that one should
multiply only the bare nucleon contribution by a normalization constant which
is
given by $\tilde Z_N = 1 - M_1^{\pi/N}$. This is of course what one gets when
eq.~(1) is expanded into a power series in the $\pi NN$ coupling constant until
order $g_{\pi NN}^2$. Numerically however, this makes a big difference. Note
that in this case the mean number of pions per nucleon, $\ave{n_\pi}$ is
identical with the first moment $M_1^\pi$. The difference between ref.
\cite{meth} and ref. \cite{ss93} (i.e. our eq.~(1))
is due to a different summation
scheme of diagrams when the dressed nucleon propagator is calculated. The
authors of ref. \cite{meth} perform a systematic expansion in the coupling
constant until quadratic order, thus taking into account only single one-loop
diagrams. In contrast, the authors of ref. \cite{ss93} iterate the meson-baryon
loops until infinite order. Very recently Holtmann showed by a comparison of
the corresponding next order corrections that this is a significantly better
approximation \cite{hh94}.
In any case, when we turn to the calculation of nuclear structure functions
a strict expansion in the $\pi NN$ coupling constant is no longer a useful
scheme. The medium modifications of the pion distribution function are mainly
due to rescattering processes and Pauli blocking. So if we restrict ourselves
to diagrams of order $g_{\pi NN}^2$ we should also drop all rescattering
diagrams. In this case we would be left with Pauli blocking as the only medium
effect which would lead to a depletion rather than an enhancement of the pion
field. On the other hand, if we sum over an infinite set of rescattering
diagrams as it is usually done \cite{etma,bbmr} we should also iterate the
$\pi N$ loops which dress the nucleon. This means we should normalize like in
eq.~(1).

In the pion excess model \cite{etma,bbmr}
it is  assumed that the quark distribution functions of the bare nucleon and
of the pion remain unchanged inside a nucleus. Thus in analogy to eq. (1) we
write
\be
   q_f^{p/A}(x) \;=\; Z_A \{ q_{f,bare}^p(x)
   \;+\; \sum_i c_i \;[\; \int_x^A \frac{dy}{y} \; f^{N/A}(y) \;
   q_{f,bare}^{N_i} (\frac{x}{y})
   \;+\;  \int_x^A \frac{dy}{y} \; f^{\pi/A}(y) \;
   q_f^{\pi_i} (\frac{x}{y})\;]\;\},
\ee
with relations analogous to eqs. (2) - (4) in order to guarantee baryon number,
charge and momentum conservation. Note that we still define the
Bjorken variable $x$ with respect to the nucleon mass. Therefore, unlike
the free nucleon case, $f^{\pi/A}$ and $f^{N/A}$ do not need to vanish
for $y \geq 1$ and in general a relation like eq.~(5) can not be correct
in a nucleus.

On the other hand in microscopic models
the upper limit $y_{max}$ for non-vanishing $f^{\pi/A}$ does not really
become equal to the nuclear mass number $A$ as we have indicated in eq.~(7)
but remains rather close to unity.
For the model presented in ref. \cite{bblw} for instance, we find
$y_{max} = 1 + \frac{k_F^2}{m_N^2}$ which is about 1.1 at nuclear matter
density (The model of refs. \cite{etma,bbmr} even leads to
$y_{max} < 1$ but that is
an artefact of the nonrelativistic kinematics.). Furthermore the shape of the
pion distribution function is not dramatically changed by nuclear effects and
the main contribution still comes from the regime $y \leq 1$. Thus
before we turn to a more realistic model we consider a schematic model where
the pion distribution function in a nucleus is simply given by
$f^{\pi/N}$ multiplied by an enhancement factor $\lambda$:
\be
   f^{\pi/A}(y) \;=\; \lambda\,f^{\pi/N}(y) \;.
\ee
For the distribution function of the recoil nucleons
$f^{N/A}(y)$ we assume that a relation analogous to eq.~(5) holds.

Within this model
we calculate EMC and Drell-Yan ratios for different values of $\lambda$.
For the
$\pi NN$ formfactor we choose a monopole parameterization with a cutoff
parameter $\Lambda_{\pi NN} = 800 MeV$.
We assume the sea quark distributions of the bare nucleons and of the pions
to be $SU(3)$-symmetric and
the proton and the neutron as well as the different pions
to be related by isospin rotations.
We parameterize the quark distributions of the bare proton in
the following simple form:
\be
   x\,q_{f,bare}^p(x) \;=\; N_f\,x^{\alpha_f}\,(1-x)^{\beta_f}  \quad .
\ee
The parameters are chosen such that the
experimental distribution functions of the dressed proton \cite{onu}
are roughly reproduced by eq.~(1).
We find $\alpha_f = 0.65$ and
$\beta_f = 3$ for valence up quarks, $\alpha_f = 0.65$ and $\beta_f = 4$
for valence down quarks and $\alpha_f = 0$, $\beta_f = 9$ and $N_f = 0.17$
for sea quarks. The normalization of the valence
quarks is of course fixed by eq.~(3).
For the pion we take the experimental quark distribution functions of
ref. \cite{opi}.

The left panel of fig. 1 shows our results for the EMC ratio $F_2^A/F_2^D$
for isospin symmetric nuclear matter. For the deuteron nuclear effects are
neglected. At small values of $x$ the ratio becomes enhanced with increasing
$\lambda$ whereas it becomes reduced at larger $x$. At $x \simeq 0.2$ the
EMC ratio remains unity independently of $\lambda$.
This behavior is easy to understand. Using $F_2(x) = x \sum e_f^2 q_f(x)$
we find for the EMC ratio at $x = 0$:
\be
    \frac{F_2^A(0)}{F_2^N(0)} \;=\; 1 \;+\; (\ave{n_\pi^A} - \ave{n_\pi^N})\,
    \frac{F_2^\pi(0)}{F_2^N(0)}  \;.
\ee
This result is quite general and does not depend on the specific assumption
eq.~(8).
For our schematic model we can also calculate
the ratio at $x = 1$. At this point the contributions of the pion and the
recoil nucleon vanish and the EMC ratio is just the ratio of the
renormalization constants:
\be
    \frac{F_2^A(1)}{F_2^N(1)} \;=\; \frac{Z_A}{Z_N} \;=\; \frac{1 -
    \ave{n_\pi^A}}{1 - \ave{n_\pi^N}} \;.
\ee
Thus with an enhanced pion field ($\ave{n_\pi^A} > \ave{n_\pi^N}$) the
EMC ratio is greater than one at $x = 0$ and less than one at $x = 1$.
So somewhere in between there must be a point $x_o$ with
$F_2^A(x_o) = F_2^N(x_o)$. In our schematic model this point is given by
by the condition
\be
     F_2^N(x_o) \;=\; F^N_{2,bare}(x_o)
\ee
independently of $\lambda$. We find $x_o \sim 0.2$. That means an enhanced pion
field leads to an EMC ratio greater than one only for $x \leq 0.2$. In order
to estimate whether this is a detectable effect we also plotted experimental
data points in fig. 1. Note that at $x \leq 0.1$ nuclear shadowing sets in
which is not included in our model. So there is only a very small window
$0.1 \leq x \leq 0.2$ where an enhancement could be seen. For realistic
parameters ($\lambda \leq 2$) this seems to be very difficult.

Here the wave function renormalization plays an important role.
As we can see from eq.~(10) the enhancement is
directly related to the number of excess pions which is given by
\be
   \ave{n_\pi^A} \,-\, \ave{n_\pi^N} \;=\;
   Z_N Z_A \,(M_1^{\pi/A} - M_1^{\pi/N}).
\ee
Thus compared with calculations without normalization factors
the effect is reduced by a factor $Z_N Z_A$. Since $Z_A < Z_N$ for an
enhanced pion field and $Z_N \simeq 0.7$ for our parameters, the additional
pion contribution is less than half of what it would be without normalization
factors. It is also crucial that the renormalization includes the pion cloud,
like in eq.~(1). If we renormalized only the bare nucleon,
like in ref. \cite{meth}, we still would get
eqs.~(10) - (12). In this case, however, the pion numbers $\ave{n_\pi}$
would be identical with the first moments $M_1^\pi$ and we would find eq.~(13)
without the factor $Z_A Z_N$.

In the right panel of fig. 1 we show the corresponding Drell-Yan ratios as a
function of $x_t$, the Bjorken variable for the target nucleus. The Bjorken
variable for the projectile was kept fixed at $x_p = 0.5$. At this large
value of $x_p$ we are mainly sensitive to the target sea which is dominated
by the pion contribution.
Therefore this process is much more sensitive to detect an enhancement of the
pion field although the effect is also strongly reduced by the normalization
factors.

After these schematic studies we want to make the model somewhat more
realistic.
The enhancement of the pion distribution function in nuclear matter has been
calculated many times \cite{etma,bbmr,bblw}. In these models the bare pion
propagator becomes dressed by nucleon hole and $\Delta$ hole excitations.
As we have mentioned before, $f^{\pi/A}(y)$ does not have to vanish for
$y \geq 1$ since a pion may carry  more than the fraction $1/A$ of the plus
momentum of the nucleus. Therefore conservation of charge and
momentum cannot be enforced by a relation like eq.~(5). This can also be seen
from more physical arguments: Since we are summing an infinite series of
particle hole polarizations we should allow the virtual photon not only to
couple to the recoil nucleon which corresponds to the first emitted pion but
also to particle states which are created by absorbing a pion, i.e. to
nucleons which have interacted with another nucleon before. The
most natural way to describe all these processes effectively
is to incorporate them into Fermi motion and nuclear binding.

We proceed as follows: Integrating eqs.~(1) and (7) over $x$ and
using charge conservation, i.e. eq.~(2) and the corresponding relation for the
nuclear case, we can eliminate the recoil nucleon distributions as well as the
distribution functions of the bare core. In this way we find expressions for
$\int dx \, q_f^{p/A}(x)$ which only depend on the distribution functions of
the (dressed) free nucleon and the pion distribution functions $f^{\pi/N}$
and $f^{\pi/A}$. By construction the integrands already satisfy the
requirements of charge conservation, but momentum is not yet conserved. In the
next step we convolute the nucleonic part with a distribution function
$f^N_{Fermi}(z)$ which describes the effects of Fermi motion and nuclear
binding. This procedure does not change the first moments of the quark
distribution functions but it changes the second moments as a function of a
binding parameter $\eta$. Thus by tuning $\eta$ to the appropriate value
we can achieve momentum conservation while keeping the charge to be conserved.

We find the following expressions for the quark distribution functions of a
nuclear proton:
\be
   q_f^{p/A}(x) =\!\int_x^A \frac{dz}{z}\,f^N_{Fermi}(z)\,q_f^p(\frac{x}{z})
   \,+ \sum_i c_i \int_x^A \frac{dy}{y}\,
   (Z_A\,f^{\pi/A}(y)\,-\,Z_N\,f^{\pi/N}(y))\, q_f^{\pi_i}(\frac{x}{y})
   \,-\, \Delta_f^p(x),
\ee
with
\ba
   \Delta_u^p(x) \,=\frac{2}{3}\, (Z_N - Z_A)
   [\int_x^A \frac{dz}{z}\,f^N_{Fermi}(z)
   (u^p(\frac{x}{z}) - u^n(\frac{x}{z})) \qquad\qquad\quad  \cr \cr
    +\, \frac{4}{3} Z_N \int_x^1 \frac{dy}{y}\,f^{\pi/N}(y)
   (u^{\pi^+}(\frac{x}{y}) - u^{\pi^o}(\frac{x}{y})) ]\,,&\cr
\ea
$\Delta_d^p(x) = - \Delta_u^p(x)$ and $\Delta_f^p(x) = 0$ in all other cases.

Except for the normalization factors the two integrals in eq.~(14) are
identical with the standard pion excess model, as discussed e.g. in ref.
\cite{bbmr}. While the first integral describes the redistribution of the
nucleons
keeping the baryon number fixed, the second integral is the contribution from
the excess pions. However, since the pion cloud around a proton carries a
positive net charge, in order to conserve the total charge
we have to correct for the extra charge of the excess pions: Due to the
emission of additional $\pi^+$ mesons the number of up quarks associated with
the baryonic part of a nuclear proton is lowered while the number of down
quarks
is enhanced. This is the origin of $\Delta_f^p$ in eq.~(14). Since this term
does not contribute to the momentum balance, only its first moment is fixed and
one can easily find other expressions which do the job as well as eq.~(15). In
practice, however, this does not really matter: Since for neutrons everything
is just the other way around, in isospin symmetric nuclear matter the
$\Delta_f$-terms completely drop out. In nuclei with a small neutron excess
they
persist, but they are strongly suppressed. In fact, for $^{56}Fe$ we find that
the influence of $\Delta_f$ can be almost neglected.
Thus the main difference to earlier calculations \cite{etma,bbmr} is
the appearance of normalization factors in eq. (14) which reduce the pion
enhancement in the same way as we have discussed for the schematic model.

Our results for the EMC and Drell-Yan ratios ($^{56}Fe$ compared with
deuterium) are shown in fig.~2. The solid lines represent the ratios obtained
from eq.~(14) imposing momentum conservation. We take the experimental quark
distributions of Owens for the pion \cite{opi} as well as for the nucleon
\cite{onu}. The nuclear
pion distribution function is calculated microscopically within the
framework of a nuclear matter RPA calculation, including $NN^{-1}$ as well
as $\Delta N^{-1}$ excitations. Short-range correlations are taken
into account by folding the potential with a correlation function
$g_c(r) = 1 - j_o(q_c r)$. In the $NN$ channel we choose $q_c= 3.94 fm^{-1}$
corresponding to the mass of the $\omega$-meson. This gives rise to a
momentum dependent Migdal parameter with the limit $g'_{NN}(q=0) = 0.57$. In
the $N\Delta$ and the $\Delta\Delta$ channel $q_c$ is tuned to reproduce
$g'_{N\Delta}(0) = g'_{\Delta\Delta}(0) = 1/3$, the classical Lorentz-Lorenz
value. For details of the interaction see ref. \cite{bblw}, although in the
present
work we do not introduce density dependent masses and coupling constants
which are discussed there.
We use the nucleon distribution function $f^N_{Fermi}(z)$
derived from a Fermi gas model using the correct relativistic normalization
\cite {llbg}.
For the Fermi momentum we take $k_F = 260 MeV$ which corresponds to $\rho =
0.87 \rho_o$, the average density of the $^{56}Fe$ nucleus.
The parameter $\eta$ is fixed by momentum conservation.
In order to compensate for the additional momentum carried by the pions
($\sim 4.5\%$) we have to choose $\eta = 0.94$.

For comparison we also show the result
obtained within the schematic model we discussed
above (dotted line). We choose $\lambda = 1.75$ which corresponds to the
enhancement of the pion field we find in the microscopic calculation.
Of course, the schematic model is too simplistic to produce the rise of the EMC
ratio seen at $x \geq 0.7$. However, for $x \leq 0.6$ and in particular for
$x \leq 0.3$ -- our region of interest -- the two models agree quite well.
This is also true more or less for the Drell-Yan ratio.
Moreover, in a more detailed
investigation (not shown in fig.~2) we find that a major part of the
differences
comes about from the fact that we do not exactly start from the same quark
distribution functions of the free nucleon. Remember, in the schematic model
we calculated $q_f^p(x)$ from eq.~(1) using the simple parameterization
eq.~(9), whereas we directly used the experimental parameterizations \cite{onu}
in the more realistic model. This is also the reason why the ratios do not
exactly agree at $x = 0$ where they are fixed by eq.~(10). So with some more
effort in finding a good parameterization for the quark distribution functions
of the bare nucleon the dotted line would come even closer to the solid curve.

The comparison shows that for this type of models the EMC ratio at small and
intermediate $x$ is to a very large extent determined by the number of excess
pions and the amount of momentum transferred from the nucleons to the pions:
The first fixes the ratio at $x = 0$ while the latter determines the slope of
the drop with increasing $x$. Details which are related to higher moments of
the nucleon and the pion distribution functions are less important. Therefore
we believe that the major conclusions we have drawn from the schematic model
remain valid for more sophisticated descriptions. In particular it will be
very difficult to produce any significant enhancement in the region
$0.1 \leq 0.3$.

As mentioned above, in order to conserve the total momentum we have to choose
the binding parameter to be $\eta = 0.94$. This
corresponds to an average separation energy of $(1 - \eta) m_N \simeq 56 MeV$
which is of course unrealistically large. In order to check the sensitivity to
this parameter we give up the requirement of momentum conservation and choose
$\eta = 0.97$, the value proposed by Li et al. \cite{llbg}. The results are
shown as the dashed dotted curves in fig.~2. Since we have not changed the
number of excess pions the ratios at $x = 0$ remain unchanged, but the EMC
ratio drops off more slowly. However, for $x \leq 0.4$, i.e. in the region
where the pion enhancement was supposed to show up, there is still no
disagreement with the data.

The momentum conserving EMC result (solid line) looks very similar to
fig.~4 of ref. \cite{bbmr}. Note however, that our short-range repulsion is
much weaker and more realistic \cite{bblw}.
Without wave function renormalization this
would lead to a much larger number of excess pions (as discussed in eq.~(13))
and consequently to a larger ratio at small $x$. In order to demonstrate
this we replace $Z_N$ and $Z_A$ in eq.~(14) by unity. The results are indicated
by the long and the short-dashed lines in fig.~2. When we
enforce momentum conservation by choosing $\eta = 0.88$ we get the results
indicated by the short-dashed line. Similar to the behavior discussed in fig.~1
the EMC ratio drops off very fast and becomes unity at $x \sim 0.2$. When we
choose the realistic parameter $\eta = 0.97$ we arrive at the long-dashed
curve. This is the only case where the EMC-data are significantly overestimated
in the whole area below $x \leq 0.3$.

To conclude, we find that EMC experiments are not sufficiently sensitive to
detect an effect due to a possibly enhanced pion field in nuclei. Within a
schematic model we show that the enhancement of the pion distribution function
must be unrealistically large to produce a significant deviation from the data
in the region $0.1 \leq x \leq 0.3$. These findings are confirmed by more
realistic calculations. Our results are based on a generalization of the
two-phase model for the free nucleon of Szczurek and Speth \cite{ss93}. In
this model the pion contribution is strongly reduced by a renormalization
constant. Other prescriptions which renormalize the bare nucleon part only
are based on a power expansion in the coupling constant and seem to be
unsuitable for nuclear matter.

In contrast to EMC experiments, Drell-Yan experiments are
sensitive to conclude that there is indeed no enhancement of the pion field in
nuclei. Although the renormalization constants reduce the discrepancy by more
than $50 \%$ the data are still overestimated in a standard RPA calculation.
This leaves room for more unconventional explanations \cite{bblw}.

So far we have restricted ourselves to nucleon and pion degrees of freedom.
Although heavier mesons are known to be more important in the intermediate
regime $x \sim 0.5$, through the normalization factors and the momentum
balance they also have some influence on the lower $x$ region. We do not
expect, however, that this will spoil our main conclusions.

\bigskip

I would like to thank G.E. Brown for many stimulating discussions and
H. Holtmann and A. Szczurek for very useful information about their model
of the nucleon.
This work was supported in part by the Feodor Lynen program of the Alexander
von Humboldt foundation. Major parts were done during my visit
to the IKP of the KFA J\"ulich
and I thank J. Speth for his hospitality.

\newpage
\pagestyle{empty}
\begin{center}
\centering{\bf{\large Figure Captions}}
\end{center}
\vspace{1.3cm}
\begin{itemize}
\item[{\bf Fig.~1}]
EMC ratio (left panel) and Drell-Yan ratio (right panel) for isospin symmetric
nuclear matter calculated within the schematic model (eq.~(8)) with $\lambda =
1$ (solid line), $\lambda = 1.5$ (dashed-dotted line), $\lambda = 2$ (dashed
line) and $\lambda = 4$ (dotted line). The data points \cite{apma,argr,adml}
are plotted in order to give an estimate for the experimental sensitivity.

\item [{\bf Fig.~2}]
EMC ratio (left panel) and Drell-Yan ratio (right panel) for $^{56}Fe$
calculated in a realistic model using eq.~(14). The solid line shows the result
for the binding parameter $\eta = 0.94$, obtained from momentum conservation,
while the dashed-dotted line is calculated with the more realistic value $\eta
= 0.97$. The dashed lines indicate the results we get in a calculation without
renormalizing the pion contribution ($Z_A = Z_N = 1$), short-dashed line:
$\eta = 0.88$, long-dashed line: $\eta = 0.97$. For comparison we also show
the result obtained within the schematic model with $\lambda = 1.75$ (dotted
line). The EMC data were taken from refs. \cite{apma} ($^{40}Ca$) and
\cite{argr} ($^{56}Fe$), the Drell-Yan data were taken from ref. \cite{adml}
($^{56}Fe$).

\end{itemize}

\end{document}